\documentclass[aps,prl,preprint,superscriptaddress,showpacs,endfloats]{revtex4-1}

\usepackage{bm}
\usepackage{amsmath,amssymb}
\usepackage{graphicx}
\usepackage{url}
\usepackage{textcase}

\def\vector#1{\mbox{\boldmath $#1$}}

\begin{document}

\title{Temperature and magnetic field dependence of the internal and lattice structures of skyrmions by off-axis electron holography}

\author{K. Shibata}
\email{kiyou.shibata@riken.jp}
\altaffiliation[Current affiliation: ]{RIKEN Center for Emergent Matter Science (CEMS), Wako 351-0198, Japan}
\affiliation{Department of Applied Physics, the University of Tokyo, Tokyo 113-8656, Japan}
\author{A. Kov{\'a}cs}
\affiliation{Ernst Ruska-Centre for Microscopy and Spectroscopy with Electrons and Peter Gr{\"u}nberg Institute, Forschungszentrum J{\"u}lich GmbH, D-52425 J{\"u}lich, Germany}
\author{N. Kanazawa}
\affiliation{Department of Applied Physics, the University of Tokyo, Tokyo 113-8656, Japan}
\author{R. E. Dunin-Borkowski}
\affiliation{Ernst Ruska-Centre for Microscopy and Spectroscopy with Electrons and Peter Gr{\"u}nberg Institute, Forschungszentrum J{\"u}lich GmbH, D-52425 J{\"u}lich, Germany}
\author{Y. Tokura}
\email{tokura@riken.jp}
\affiliation{Department of Applied Physics, the University of Tokyo, Tokyo 113-8656, Japan}
\affiliation{RIKEN Center for Emergent Matter Science (CEMS), Wako 351-0198, Japan}

\date{\today}

\begin{abstract}
The internal and lattice structures of magnetic skyrmions in \textit{B}20-type FeGe are investigated using off-axis electron holography.
The temperature, magnetic field and angular dependence of the magnetic moments of individual skyrmions are analyzed.
Whereas the internal skyrmion shape is found to vary with magnetic field, the inter-skyrmion distance remains almost unchanged in the lattice phase.
The amplitude of the local magnetic moment is found to depend on temperature, while the skyrmion shape does not.
Deviations from a circular to a hexagonal skyrmion structure are observed in the lattice phase.
\end{abstract}

\pacs{41.20.Gz,68.37.Lp,75.25.-j}

\maketitle

Skyrmions are nanoscale vortex-like spin objects \cite{Bogdanov1994,Roessler2011,Nagaosa2013} that are stabilized in chiral crystals \cite{Muhlbauer2009,Yu2010} and bilayer films \cite{Heinze2011,Romming2013,Romming2015} due to the Dzyaloshinskii-Moriya interaction \cite{Dzyaloshinsky1958, Moriya1960}.
Skyrmions and skyrmion lattices (SkLs) \cite{Muhlbauer2009,Yu2010}, are attracting considerable attention as a result of their novel electrodynamic properties and potential applications \cite{Nagaosa2013,Sampaio2013}.
Although surface-sensitive spin polarized scanning tunneling microscopy has been used to study bilayer PdFe/Ir(111), revealing the structures of atomic-scale SkLs \cite{Heinze2011} and the magnetic field dependence of isolated skyrmions \cite{Romming2015}, detailed information about the SkL structure in chiral crystals remains elusive.
Skyrmions and SkLs in chiral crystals can be regarded as spatially-localized particle-like objects that can assemble together and are topologically protected \cite{Bogdanov1994,Nagaosa2013}, as confirmed by real-space observations \cite{Yu2010,Romming2013,Romming2015,Milde2013}.
SkLs have been described properly as multiple-$Q$ states in small-angle neutron scattering (SANS) studies \cite{Adams2011,Grigoriev2014}.
However, the detailed real-space investigation of the dependence of skyrmion structure on temperature $T$ and external magnetic field $B_\mathrm{ext}$ is still lacking.

In this Letter, we investigate the $T$- and $B_\mathrm{ext}$- dependence of the internal and lattice structures of skyrmions using off-axis electron holography in the transmission electron microscope (TEM).
The technique can be used to provide real-space measurements of in-plane magnetic induction projected onto a plane perpendicular to the incident electron beam direction ($\vector{k}_\mathrm{e}$) \cite{Lichte2008}.
Previously, Park \textit{et al.} observed a SkL in Fe$_{0.5}$Co$_{0.5}$Si using off-axis electron holography and discussed its three-dimensional structure on the basis of the thickness-dependence of the recorded electron phase shift ($\phi$) \cite{Park2014}.
In contrast, here we focus on the spatial distribution of $\phi$ and discuss the dependence of internal skyrmion and SkL structure on $T$ and $B_\mathrm{ext}$.
We examine a thin plate of \textit{B}20-type FeGe, which has suitable physical properties for electron holography measurements:
a high critical temperature $T_\mathrm{C}$ ($278\ \mathrm{K}$),
a relatively long periodicity of the helical magnetic order ($70\ \mathrm{nm}$),
a large magnetization ($1\ \mu_\mathrm{B}$/FeGe)
and a wide SkL phase region in the $T-B_\mathrm{ext}$ plane for thin samples \cite{Yu2011}.

\begin{figure}
\includegraphics{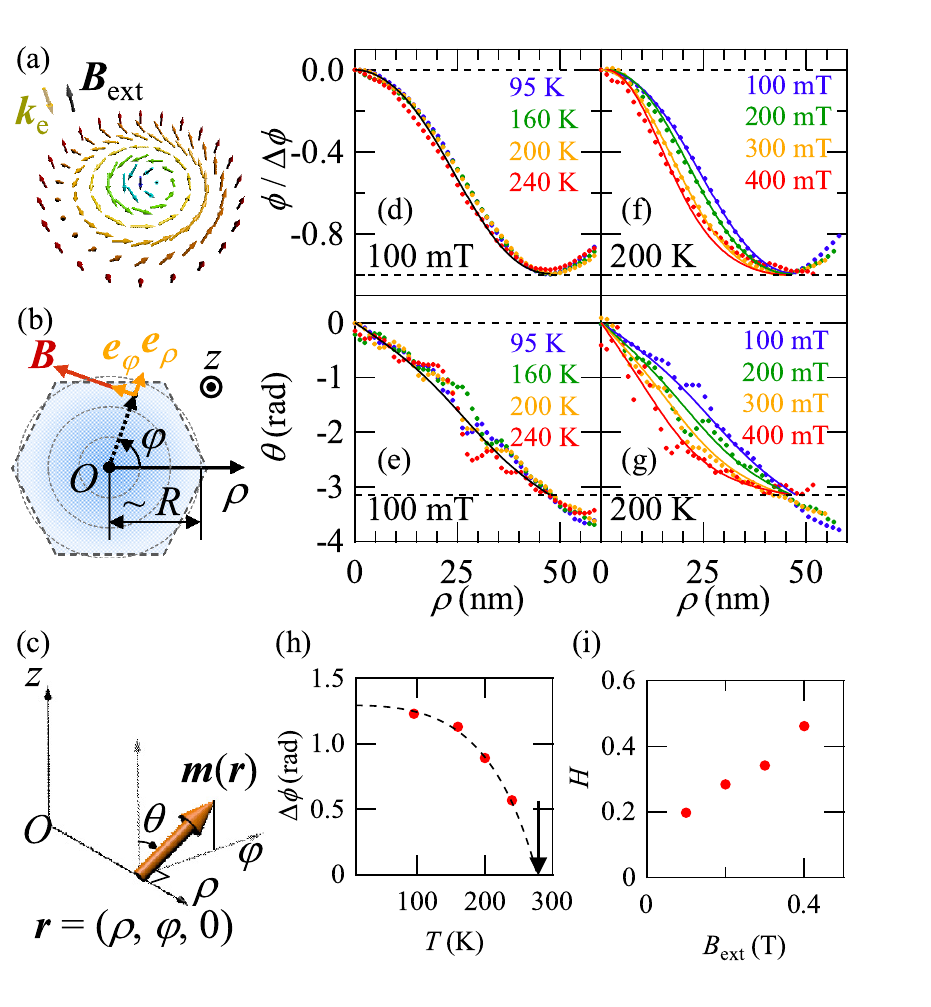}
\caption{
Analysis of the dependence of internal skyrmion structure on $T$ and $B_\mathrm{ext}$.
(a) Schematic diagram of the magnetic configuration ($\vector{m}$) of a skyrmion and the experimental configuration.
$\vector{B}_\mathrm{ext}$ and $\vector{k}_\mathrm{e}$ are the external magnetic field direction and the incident electron beam direction, respectively.
(b) Polar coordinate system ($\rho,\varphi, z$) used to describe the skyrmion structure.
(c) Definition of spin rotation angle $\theta$.
(d) $T$-dependence of the normalized phase distribution plotted as a function of distance ($\rho$) from the center of each skyrmion.
(e) Spin rotation angle $\theta$ plotted as a function of $\rho$.
(f) $B_\mathrm{ext}$-dependence of the normalized phase distribution plotted as a function of $\rho$.
(g) $B_\mathrm{ext}$-dependence of $\theta$ plotted as a function of $\rho$.
(h) Normalization factor $\Delta\phi$ for the phase curves shown in (d) (see text for details), plotted as a function of $T$.
The dashed line is a guide to the eye. The vertical arrow indicates the magnetic transition temperature.
(i) Fitting parameter $H$ (for definition see Eq.~(\ref{eq:diff_eq_bogdanov}) in the text) for the curves shown in (d)-(g), plotted as a function of $B_\mathrm{ext}$.
\label{fig:3}
}
\end{figure}

Single crystals of \textit{B}20-type FeGe were synthesized using a chemical vapor transport method \cite{Richardson1967}.
A thin plate-like sample of FeGe (110) with a thickness of approximately $120\ \mathrm{nm}$ was prepared for off-axis electron holography using a focused ion beam instrument (NB-5000, Hitachi).
Off-axis electron holograms were recorded using an accelerating voltage of $300\ \mathrm{kV}$ in an FEI Titan TEM equipped with a field emission electron gun, multiple electron biprisms and an objective lens spherical aberration corrector.
The sample temperature was controlled using a liquid N$_2$ cooling holder (Gatan model 636).
An external magnetic field $\vector{B}_\mathrm{ext}$ was applied parallel to the incident electron beam direction (\emph{i.e.}, perpendicular to the plane of the thin sample) using the magnetic field of the partially-excited conventional microscope objective lens, thereby allowing skyrmions to be stabilized \cite{Yu2010}.
Real-space phase images were reconstructed from recorded holograms using HoloWorks software (Gatan).
The structures of the observed skyrmions were analyzed in cylindrical coordinates: $\vector{r}=(\rho \sin \varphi, \rho \cos \varphi, z)$,
where the origin ($\vector{r} = \vector{0}$) is taken at the center of each skyrmion and the $z$ axis is perpendicular to the skyrmion plane (see Figs.~\ref{fig:3}(b) and \ref{fig:3}(c) below).
When an incident electron travels along the positive $z$ direction, it experiences a phase shift that can be described using the equation
\begin{equation}
	\phi(\rho, \varphi)=\frac{e}{\hbar\nu}\int_{z=-\infty}^{+\infty} V(\rho, \varphi, z) \mathrm{d}z -\frac{e}{\hbar}\int_{z=-\infty}^{+\infty} \vector{A}(\rho, \varphi, z) \cdot \vector{e}_z \mathrm{d}z,
	\label{eq:phase}
\end{equation}
where $V$ is the mean inner electrostatic potential of the sample, $\nu$ is the electron velocity, $\vector{A}$ is the magnetic vector potential, $\hbar$ is the reduced Planck constant, $e$ is the elementary electric charge, and $\vector{e}_i$ is a fundamental unit vector in the coordinate system ($i=\rho,\varphi,z$)\cite{Lichte2008}.
The first term is the electrostatic contribution to the phase shift, while the second term is the magnetic contribution.

\begin{figure}
\includegraphics{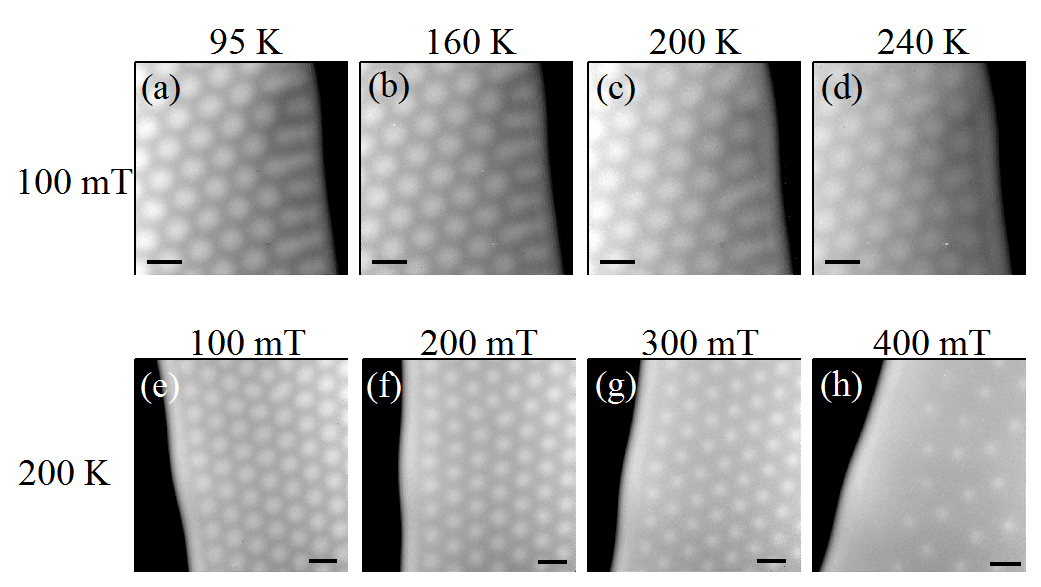}
\caption{
Dependence on temperature $T$ and applied magnetic field $B_\mathrm{ext}$ of the phase shift ($\phi$) recorded from skyrmions in a thin sample of FeGe using off-axis electron holography.
(a)-(d) Phase shift recorded in a magnetic field of $100\ \mathrm{mT}$ at a temperature of (a) $95\ \mathrm{K}$, (b) $160\ \mathrm{K}$, (c) $200\ \mathrm{K}$, and (d) $240\ \mathrm{K}$.
(e)-(h) Phase shift recorded at a temperature of $200\ \mathrm{K}$ in a magnetic field of (e) $100\ \mathrm{mT}$, (f) $200\ \mathrm{mT}$, (g) $300\ \mathrm{mT}$, and (h) $400\ \mathrm{mT}$.
The magnetic field was always applied perpendicular to the plane of the sample ($\simeq 100\ \mathrm{nm}$ thick).
All scale bars are $100\ \mathrm{nm}$.
\label{fig:1}
}
\end{figure}
Figure~\ref{fig:1}(a) shows a representative reconstructed phase image recorded at $95\ \mathrm{K}$ and $100\ \mathrm{mT}$ after field cooling at $100\ \mathrm{mT}$.
The hexagonally-arranged peaks in the recorded phase image, each of which corresponds to a skyrmion that has a counter-clockwise in-plane winding of its magnetic moments \cite{Park2014}, confirms the formation of a SkL even at $95\ \mathrm{K}$ ($\ll T_\mathrm{C}\sim 278\ \mathrm{K}$) \cite{Milde2013}.
Figures~\ref{fig:1}(b)-(d) show the distribution of the phase $\phi$ after heating from $95\ \mathrm{K}$ in the presence of a $100\ \mathrm{mT}$ magnetic field.
The dependence of $\phi$ on $B_\mathrm{ext}$ was also investigated. After a SkL had formed at $200\ \mathrm{K}$ in $100\ \mathrm{mT}$ by field cooling, $B_\mathrm{ext}$ was increased.
Figures~\ref{fig:1}(e)-(h) show the distribution of $\phi$ measured at different values of $B_\mathrm{ext}$ at $200\ \mathrm{K}$.
The SkL survives up to approximately $350\ \mathrm{mT}$. The application of a $400\ \mathrm{mT}$ field then annihilates some skyrmions, while some remain (Fig.~\ref{fig:1}(h)).
At $450\ \mathrm{mT}$ (not shown), no phase peaks are observed, all skyrmions are annihilated and a ferromagnetic state is realized.

\begin{figure}
\includegraphics{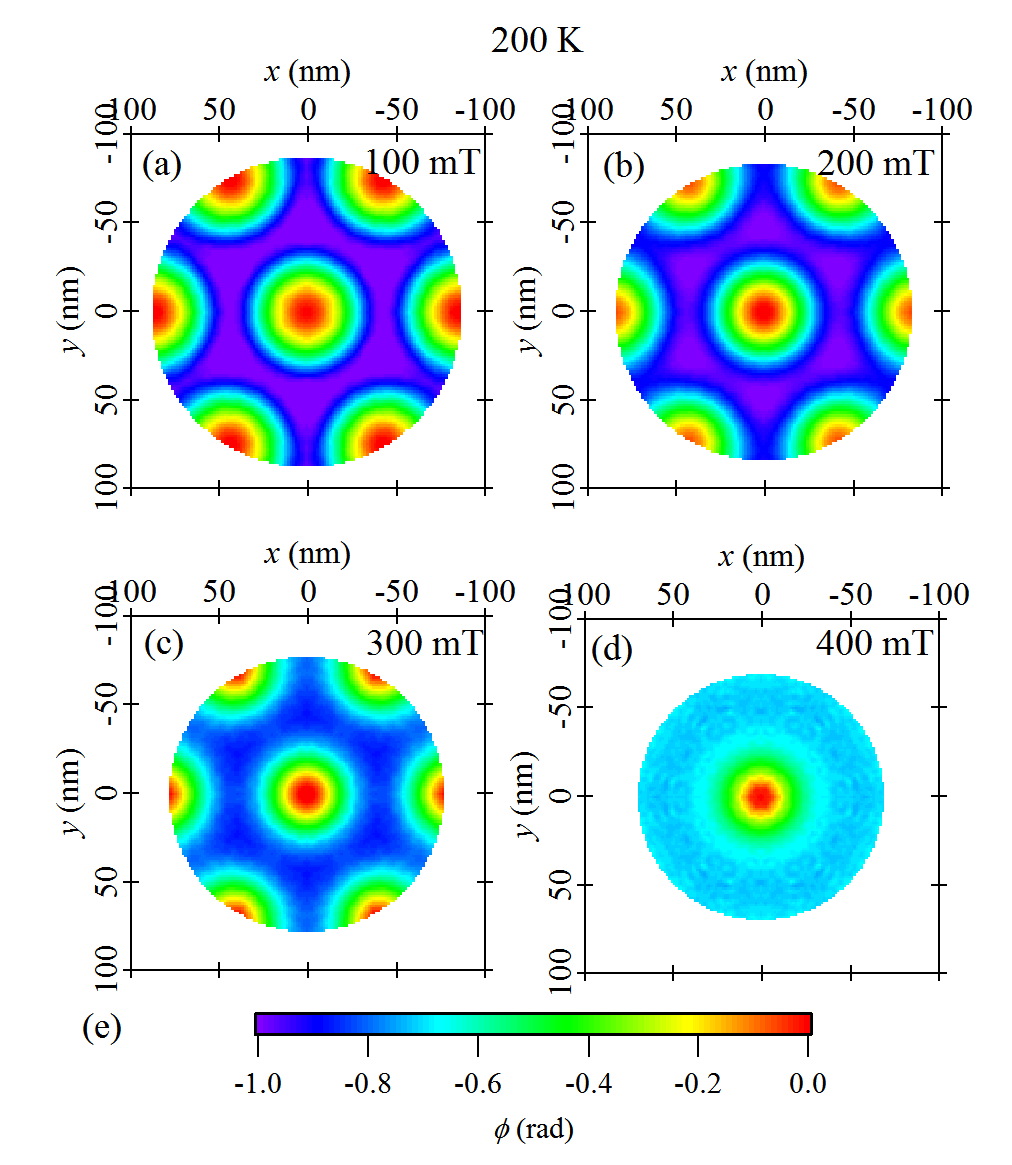}
\caption{
Symmetrized maps of the experimentally measured phase shift $\phi$ recorded from skyrmions at $200\ \mathrm{K}$ in an applied magnetic field of (a) $100\ \mathrm{mT}$, (b) $200\ \mathrm{mT}$, (c) $300\ \mathrm{mT}$ and (d) $400\ \mathrm{mT}$, averaged over 22, 33, 39 and 12 skyrmions, respectively, making use of the six-fold and mirror symmetry of the lattice.
The origin of the phase shift $\phi$ is set to 0 at the center of each skyrmion.
(e) Color scale of the phase shift in (a)-(d).
\label{fig:2}
}
\end{figure}
Assuming a uniform sample thickness $t$ and mean inner potential $V$, \textit{i.e.}, a flat and homogeneous sample, the first term in Eq.~(\ref{eq:phase}) takes the form of a uniform phase offset and can be ignored.
In order to reduce the influence of statistical noise, the recorded phase was averaged over different skyrmion sites and symmetrized by taking into account the six-fold and mirror symmetry of the lattice.
Figure~\ref{fig:2} shows symmetrized experimental maps of $\phi(\rho,\varphi)$ for different values of $B_\mathrm{ext}$ at $200\ \mathrm{K}$.

We analyzed the magnetic configuration of each unit cell of the SkL using the symmetrized $\phi(\rho,\varphi)$ maps.
Figure~\ref{fig:3}(a) shows a schematic magnetic configuration of a skyrmion.
The magnetic moment vectors $\vector{m}$ at the core and edge are anti-parallel and parallel to the applied magnetic field $\vector{B}_\mathrm{ext}=B_\mathrm{ext}\vector{e}_z$, respectively \cite{Muhlbauer2009,Yu2010}.
In the intermediate region, $\vector{m}$ rotates from the core to the edge in a unique rotational sense, which is right-handed in this case.
We assume a cylinder-like structure, in which $\vector{m}(\vector{r})$ is homogeneous along $\vector{B}_\mathrm{ext} || \vector{e}_z$.
As far as the unit cell of the SkL is concerned, it is then convenient to describe the structure as a function of distance from the center of a skyrmion ($\rho$), on the assumption that the skyrmion can be treated as an almost axially symmetrical object.
In order to compare the shapes of skyrmions determined from different phase images, each radial phase profile $\phi(\rho)$ was normalized, so that the difference ($\Delta\phi$) between the peak and the dip was identical, as shown in Fig.~\ref{fig:3}(d) for different values of $T$ and $B_\mathrm{ext}= 100\ \mathrm{mT}$.
The consistency between the curves indicates that there is no significant dependence of skyrmion structure on $T$.
Figure~\ref{fig:3}(h) shows the dependence of the normalization factor $\Delta\phi$ on $T$.
The decrease in $\Delta\phi$ with increasing $T$ is attributed to the decrease in the effective magnetic moment due to thermal fluctuations. The $T$-dependent variation of $\Delta\phi$ appears to be consistent with that of the magnitude of the ordered moment \cite{Lebech1989}.
Figure~\ref{fig:3}(f) shows normalized profiles of $\phi(\rho)$ plotted for different values of $B_\mathrm{ext}$ at $200\ \mathrm{K}$.
In contrast to the absence of a $T$-dependence, the $\phi(\rho)$ profiles vary considerably with $B_\mathrm{ext}$.

The spatial distribution of the direction of $\vector{m}(\vector{r})$ was used as an intuitively-understandable quantity to interpret the magnetic configurations from the $\phi(\rho)$ curves.
On the assumption that the thickness $t$ and mean inner potential $V$ are uniform in the sample and that there are no strong magnetic fringing fields, $\nabla\phi$ is proportional to the projected in-plane magnetic induction $\vector{B}$, which is in turn approximately proportional to the projected in-plane magnetization \cite{Dunin-Borkowski1998}.
In the present coordinate system, we assume that the relation $\nabla\phi(\vector{r}) = - \frac{et}{\hbar}\vector{B}(\vector{r})\times (-\vector{e}_z)$ holds. In particular, we make use of the relation $\frac{\partial\phi}{\partial\rho}=\vector{e}_{\rho} \cdot \nabla\phi(\vector{r})= \frac{et}{\hbar}\vector{B}(\vector{r})\cdot \vector{e}_\varphi \propto t\vector{m}(\vector{r}) \cdot \vector{e}_{\varphi}(\vector{r})$ along the radial ($\rho$) direction.
Assuming a fixed magnetization amplitude $M$ and a fixed Bloch-type magnetic helicity \cite{Nagaosa2013} in FeGe \cite{Uchida2008,Yu2011}, $\vector{m}(\vector{r})$ can be described in the form $\vector{m} = M (\sin \theta \cos \varphi, \sin \theta \sin \varphi, \cos \theta)$, where $\theta (\rho)$ is the spin rotation angle, as defined in Fig.~\ref{fig:3}(c).
Then, $\theta(\rho)$ is given by the expression
\begin{equation}
	\theta(\rho) = \sin^{-1} \left(\frac{\vector{m}(\rho) \cdot\vector{e}_\varphi}{M}\right) = \sin ^{-1}\left(\frac{\partial\phi}{\partial\rho}/\left.{\frac{\partial\phi}{\partial\rho}}\right|_{\rm max}\right),
	\label{eq:theta}
\end{equation}
where $\left.{\frac{\partial\phi}{\partial\rho}}\right|_{\rm max}$ is the maximum value of $\left|{\frac{\partial\phi}{\partial\rho}}\right|$.
Figures~\ref{fig:3}(e) and \ref{fig:3}(g) show the measured $T$- and $B_\mathrm{ext}$- dependence of $\theta$ plotted as a function of $\rho$, calculated from Figs.~\ref{fig:3}(d) and \ref{fig:3}(e), respectively, using Eq.~(\ref{eq:theta}).
The value of $\rho$ at which $\vector{m}$ shows an in-plane direction ($\theta=\frac{\pi}{2}$) becomes smaller with increasing $B_\mathrm{ext}$.
This behavior can be understood intuitively by considering an increasing contribution of $\vector{m}(\vector{r})$ parallel to $\vector{B}_\mathrm{ext}$ at large values of $B_\mathrm{ext}$ due to a gain in Zeeman energy.
Meanwhile, the inter-skyrmion distance $a_\mathrm{Sk}$, which is double the value of $\rho$ at which $\theta = -\pi$, is fixed at approximately $90\ \mathrm{nm}$ and is unaffected by $B_\mathrm{ext}$, as shown in Fig.~\ref{fig:3}(g).
This observation indicates that the competition between the exchange energy and the Dzyaloshinskii-Moriya interaction determines the magnetic periodicity, as predicted by Bogdanov and Hubert \cite{Bogdanov1994}.

We fitted the experimental data to a theoretical model proposed by Bogdanov and Hubert \cite{Bogdanov1994}, who approximated a skyrmion configuration in a SkL in terms of a circular object and determined the $\theta(\rho)$ profile as a solution of the differential equation
\begin{equation}
	\frac{d^2\theta}{d\rho^2} + \frac{1}{\rho^2}\sin \theta \cos \theta + \frac{\sin^2 \theta}{\rho}-\frac{1}{2}H \sin \theta -\tilde{\beta} \sin \theta \cos \theta = 0,
	\label{eq:diff_eq_bogdanov}
\end{equation}
where $H$ is the normalized magnetic field and $\tilde{\beta}$ is an anisotropy constant along the $z$ direction.
The boundary condition for Eq.~(\ref{eq:diff_eq_bogdanov}) is chosen according to the surrounding condition of the skyrmion:
$\theta(\rho=0)=\pi, \theta(\rho= \infty)=0$ for an isolated skyrmion ($400\ \mathrm{mT}$),
while $\theta(\rho=0)=\pi, \theta(\rho= R)=0$ for a skyrmion in a SkL state ($100\ \mathrm{mT}$, $200\ \mathrm{mT}$, and $300\ \mathrm{mT}$).
Considering the relatively weak crystalline anisotropy of FeGe \cite{Uchida2008}, we fitted the $\theta(\rho)$ profiles using values of $\tilde{\beta}=0$ and $R=8$, while adjusting the value of $H$ \cite{Bogdanov1994}.
The calculated curves, which are shown as solid lines in Figs.~\ref{fig:3}(f) and \ref{fig:3}(g), provide a good fit to the data deduced from the $\phi$ maps.
The fitting parameter $H$ used in Eq.~(\ref{eq:diff_eq_bogdanov}) is plotted in Fig.~\ref{fig:3}(i) as a function of the experimentally applied magnetic field $B_\mathrm{ext}$.
The high degree of linearity also confirms the validity of the Bogdanov-Hubert model for the effect of applied magnetic field on internal skyrmion structure.

\begin{figure}
\includegraphics{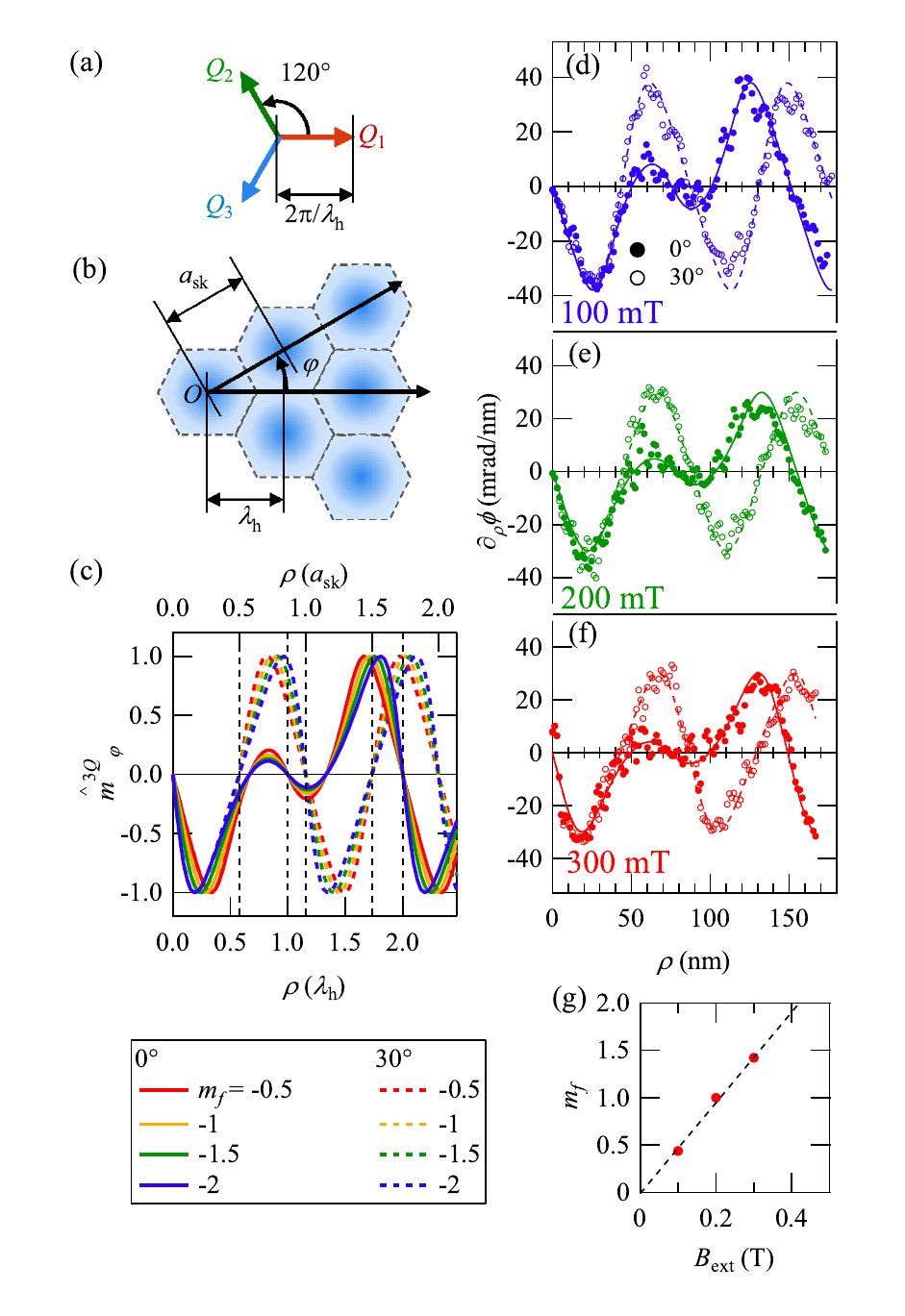}
\caption{
Analysis of the structure of a SkL.
(a) Three-$Q$ configuration of a SkL.
(b) Definition of two typical directions in a SkL.
(c) Line profiles of ${\hat{m}^{3Q}}_\varphi$, the tangential component of the normalized magnetic moment defined in Eq.~(\ref{eq:normalized_3Q_model}), plotted as a function of $\rho$ for different values of $m_f$.
(d)-(f) $\rho$-derivative of the phase shift $\phi$ for  values of $B_\mathrm{ext}$ of (d) $100\ \mathrm{mT}$, (e) $200\ \mathrm{mT}$ and (f) $300\ \mathrm{mT}$.
(g) $B_\mathrm{ext}$-dependence of the fitting parameter $m_f$ (see Eq.~(\ref{eq:3Q_model}) in the text) used to determine the calculated curves in (d)-(f).
The dashed line is a guide to the eye.
\label{fig:4}
}
\end{figure}
A further interesting question is related to the deformation of the SkL structure when the magnetic field is changed.
The Bogdanov-Hubert model described above is useful for describing the internal structure of a skyrmion. However, it cannot be used to describe a hexagonal crystal structure, since it approximates a skyrmion as an axially symmetric object.
In order to explain neutron scattering results, M{\"u}hlbauer \textit{et al.} proposed an approximate magnetic configuration in terms of a composition of three helical magnetic configurations and a component along the magnetic field direction \cite{Muhlbauer2009}, in the form
\begin{equation}
	\vector{m}^{3Q}(\vector{r}) = \sum_{i=1}^{3} \vector{m}^{h}_{\vector{Q}_i}(\vector{r})+{\bm m_f},
	\label{eq:3Q_model}
\end{equation}
where $\vector{Q}_i$ ($i = 1,2,3$) are helical modulation wave vectors that satisfy the relations $\sum_{i=1,2,3} \vector{Q}_i=0$ and $\vector{Q}_i\cdot\vector{B}_\mathrm{ext}=0$, as shown in Fig.~\ref{fig:4}(a), while $\vector{m}^{h}_{\vector{Q}_i}$ is the helical spin order given by the expression
\begin{equation}
	\vector{m}^{h}_{\vector{Q}_i}(\vector{r})=A \left[ \vector{n}_{i1}\cos(\vector{Q}_i \cdot \vector{r})+\vector{n}_{i2}\sin(\vector{Q}_i \cdot \vector{r}) \right],
\end{equation}
where the vectors $\vector{n}_{i1}$ and $\vector{n}_{i2}$ define the spin-rotation plane and are perpendicular to $\vector{Q}_i$.
This simple triple-$\vector{Q}$ model describes the hexagonal lattice periodicity. However, it has the drawbacks that the amplitude of the local magnetic moment $|\vector{m}^{3Q}(\vector{r})|$ is not fixed spatially and the in-plane component of $\vector{m}^{3Q}(\vector{r})$ is independent of the $\vector{B}_\mathrm{ext}$-parallel component $\vector{m}_f$.
Figures~\ref{fig:3}(f) and \ref{fig:3}(g) show that the in-plane magnetic configuration depends on $B_\mathrm{ext}$, which cannot be accounted for by the triple-$\vector{Q}$ model.
An alternative model, which has been discussed by Adams \textit{et al.} \cite{Adams2011}, describes the magnetic structure using a higher order modulation. However, it  results in a non-uniform amplitude of the magnetic moment.

We evaluated the SkL periodicity and its dependence on $B_\mathrm{ext}$ using a phenomenological model, which approximates the SkL configuration by making use of a normalized configuration of $\vector{m}^{3Q}(\vector{r})$ in Eq.~(\ref{eq:3Q_model}), in the form
\begin{equation}
	{\hat{\vector{m}}^{3Q}}(\vector{r}) = \frac{\vector{m}^{3Q}(\vector{r})}{|\vector{m}^{3Q}(\vector{r})|}.
	\label{eq:normalized_3Q_model}
\end{equation}
In this model, the amplitude of the magnetic moment is fixed to unity, while the distribution of the in-plane component of the magnetic moment depends on $m_f$.
We compared the spatial profile of $m_\varphi(\rho)=\vector{m}(\rho)\cdot\vector{e}_\varphi$ calculated from the $\phi$ maps and ${\hat{m}^{3Q}}_\varphi (\rho)={\hat{\vector{m}}^{3Q}}_\varphi (\rho)\cdot \vector{e}_\varphi$ along two representative directions, $0^\circ$ and $30^\circ$, in the hexagonal SkL shown in Fig.~\ref{fig:4}(b).
Figure~\ref{fig:4}(c) shows the in-plane magnetic moment along these two directions calculated for various values of $m_f$.
Along the $\varphi=0^\circ$ direction, the magnetic structure has a periodicity of $2\lambda_h=2\frac{2\pi}{\left|\vector{Q}_i\right|}$, while along the $\varphi=30^\circ$ direction it has the periodicity of the SkL constant $a_\mathrm{Sk}=\frac{2}{\sqrt{3}}\lambda_h$.
Figures~\ref{fig:4}(d)-(f) show $\frac{\partial \phi}{\partial \rho}$ calculated from the observed phase maps and the model.
The hexagonal periodicity and $B_\mathrm{ext}$-dependent modulation are reproduced by the model based on the normalized moment.
Furthermore, along the $\varphi=0^\circ$ and $\varphi=30^\circ$ directions, the curve crosses 0 at different values of $\rho$ ($45\ \mathrm{nm}$ and $50\ \mathrm{nm}$, respectively).
This difference corresponds to the deformation of a skyrmion in the unit cell due to the presence of surrounding skyrmions in the SkL state.
Figure~\ref{fig:4}(g) shows the magnetic field dependence of the structural fitting parameter $\vector{m}_f$ used for fitting the experimental data in Figs.~\ref{fig:4} (d)-(f).
The linear increase of $m_f$ with $B_\mathrm{ext}$ appears to be reasonable.

In summary, we have investigated the internal and lattice structures of skyrmions in a thin sample of FeGe using off-axis electron holography.
Our electron holography measurement reveals a dependence of the local magnetic moment on $T$, as well as an internal shrinkage of the skyrmions with increasing $B_\mathrm{ext}$.
The in-plane profile of the magnetic moment in a unit cell of the SkL measured from an electron holographic phase image can be fitted well using the Bogdanov-Hubert model.
The in-plane profile of the magnetic moment also reveals that each skyrmion in the lattice is deformed from a circular to a hexagonal shape.
We used a phenomenological normalized-moment model to describe the change in the in-plane profile of the magnetic moment with applied magnetic field strength.
Our results provide a firm basis for understanding the structures of skyrmions and SkLs.

\begin{acknowledgments}
The authors would like to thank X. Z. Yu, D. Morikawa, T. Kurumaji, Y. Okamura, J. Caron and Z.-A. Li for helpful discussions.
This study was supported by the Grant-in-Aid for Scientific Research (Grant No.~24224009 and 14J09358) from the MEXT and by the Funding Program for World-Leading Innovative R\&D on Science and Technology (FIRST Program).
The research leading to these results received funding from the European Research Council under the European Union's Seventh Framework Programme (FP7/2007-2013) and under ERC grant agreement number 320832.
K.S. was supported by the Japan Society for the Promotion of Science through Program for Leading Graduate Schools (MERIT).
\end{acknowledgments}


%
\end{document}